# BundleFusion: Real-time Globally Consistent 3D Reconstruction using On-the-fly Surface Re-integration

ANGELA DAI, Stanford University
MATTHIAS NIESSNER, Stanford University
MICHAEL ZOLLHÖFER, Max-Planck-Institute for Informatics
SHAHRAM IZADI, Microsoft Research
CHRISTIAN THEOBALT, Max-Planck-Institute for Informatics

Real-time, high-quality, 3D scanning of large-scale scenes is key to mixed reality and robotic applications. However, scalability brings challenges of drift in pose estimation, introducing significant errors in the accumulated model. Approaches often require hours of offline processing to globally correct model errors. Recent online methods demonstrate compelling results, but suffer from: (1) needing minutes to perform online correction, preventing true real-time use; (2) brittle frame-to-frame (or frame-to-model) pose estimation resulting in many tracking failures; or (3) supporting only unstructured point-based representations, which limit scan quality and applicability. We systematically address these issues with a novel, real-time, end-to-end reconstruction framework. At its core is a robust pose estimation strategy, optimizing per frame for a global set of camera poses by considering the complete history of RGB-D input with an efficient hierarchical approach. We remove the heavy reliance on temporal tracking, and continually localize to the globally optimized frames instead. We contribute a parallelizable optimization framework, which employs correspondences based on sparse features and dense geometric and photometric matching. Our approach estimates globally optimized (i.e., bundle adjusted) poses in real-time, supports robust tracking with recovery from gross tracking failures (i.e., relocalization), and re-estimates the 3D model in real-time to ensure global consistency; all within a single framework. Our approach outperforms state-of-the-art online systems with quality on par to offline methods, but with unprecedented speed and scan completeness. Our framework leads to a comprehensive online scanning solution for large indoor environments, enabling ease of use and high-quality results[1].

CCS Concepts: •**Computing methodologies →Computer graphics;** *Shape modeling*; Mesh geometry models;

General Terms: RGB-D, scan, real-time, global consistency, scalable

## 1 INTRODUCTION

We are seeing a renaissance in 3D scanning, fueled both by applications such as fabrication, augmented and virtual reality, gaming and robotics, and by the ubiquity of RGB-D cameras, now even available in consumer-grade mobile devices. This has opened up the need for *real-time* scanning at *scale*. Here, the user or robot must scan an entire room (or several spaces) in real-time, with instantaneous and continual integration of the accumulated 3D model into the desired application, whether that is robot navigation, mapping the physical

---

[1] Our source code and all reconstruction results are publicly available: http://graphics.stanford.edu/projects/bundlefusion/

world into the virtual, or providing immediate user feedback during scanning.

However, despite the plethora of reconstruction systems, we have yet to see a single holistic solution for the problem of real-time 3D reconstruction at scale that makes scanning easily accessible to untrained users. This is due to the many requirements that such a solution needs to fulfill:

**High-quality surface modeling.** We need a single textured and noise-free 3D model of the scene, consumable by standard graphics applications. This requires a high-quality representation that can model *continuous surfaces* rather than discrete points.

**Scalability.** For mixed reality and robot navigation scenarios, we need to acquire models of entire rooms or several large spaces. Our underlying representation therefore must handle both small- and large-scale scanning while preserving both global structure and maintaining high local accuracy.

**Global model consistency.** With scale comes the need to correct pose drift and estimation errors, and the subsequent distortions in the acquired 3D model. This correction is particularly challenging at real-time rates, but is key for allowing online revisiting of previously scanned areas or loop closure during actual use.

**Robust camera tracking.** Apart from incremental errors, camera tracking can also fail in featureless regions. In order to recover, we require the ability to relocalize. Many existing approaches rely heavily on proximity to the previous frame, limiting fast camera motion and recovery from tracking failure. Instead, we need to (re)localize in a robust manner without relying on temporal coherence.

**On-the-fly model updates.** In addition to robust tracking, input data needs to be integrated to a 3D representation and interactively visualized. The challenge is to update the model after data has been integrated, in accordance with the newest pose estimates.

**Real-time rates.** The ability to react to instantaneous feedback is crucial to 3D scanning and key to obtaining high-quality results. The real-time capability of a 3D scanning method is fundamental to AR/VR and robotics applications.

Researchers have studied specific parts of this problem, but to date there is no single approach to tackle all of these requirements in real time. This is the very aim of this paper, to systematically address *all* these requirements in a single, end-to-end real-time reconstruction framework. At the core of our method is a robust pose estimation strategy, which globally optimizes for the camera trajectory per frame, considering the complete history of RGB-D input in an efficient *local-to-global* hierarchical optimization framework. Since we globally correlate each RGB-D frame, loop closure is handled implicitly and continuously, removing the need for any explicit loop closure detection. This enables our method to be extremely robust to tracking failures, with tracking far less brittle than existing frame-to-frame or frame-to-model RGB-D approaches. If tracking failures occur, our framework instantaneously relocalizes in a globally consistent manner, even when scanning is interrupted and restarted from a completely different viewpoint. Areas can also be revisited multiple times without problem, and reconstruction quality continuously improves. This allows for a robust scanning experience, where even novice users can perform large-scale scans without failure.

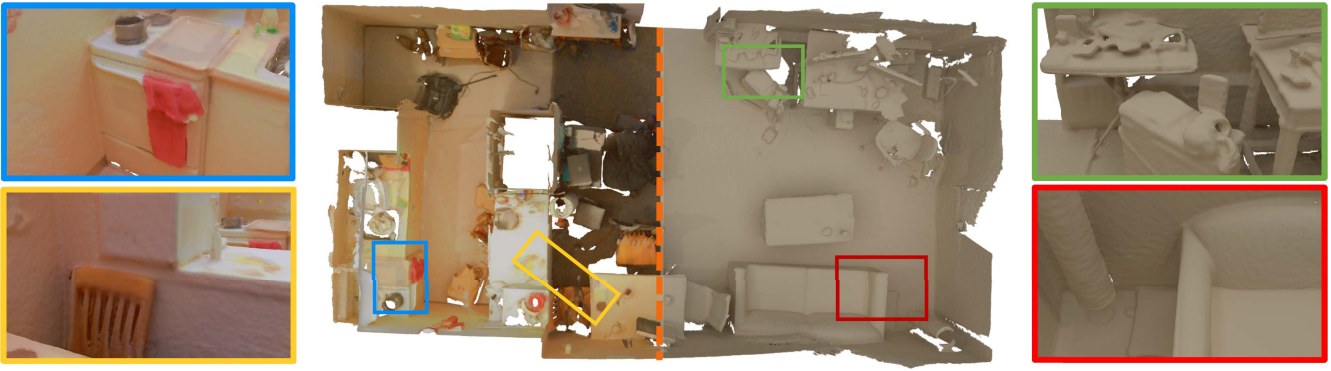

Fig. 1. Our novel real-time 3D reconstruction approach solves for global pose alignment and obtains dense volumetric reconstructions at a level of quality and completeness that was previously only attainable with offline approaches.

Key to our work is a new fully parallelizable *sparse-then-dense* global pose optimization framework: sparse RGB features are used for coarse global pose estimation, ensuring proposals fall within the basin of convergence of the following dense step, which considers both photometric and geometric consistency for fine-scale alignment. Thus, we maintain global structure with implicit loop closures while achieving high local reconstruction accuracy. To achieve the corresponding model correction, we extend a scalable variant of real-time volumetric fusion [37], but importantly support model updates based on refined poses from our global optimization. Thus, we can correct errors in the 3D model in real time and revisit existing scanned areas. We demonstrate how our approach outperforms current state-of-the-art online systems at unprecedented speed and scan completeness, and even surpasses the accuracy and robustness of offline methods in many scenarios. This leads to a comprehensive real-time scanning solution for large indoor environments, that requires little expertise to operate, making 3D scanning easily accessible to the masses.

In summary, the main contributions of our work are as follows:

(1) A novel, real-time global pose alignment framework which considers the complete history of input frames, removing the brittle and imprecise nature of temporal tracking approaches, while achieving scalability by a rapid hierarchical decomposition of the problem by using a *local-to-global optimization* strategy.

(2) A *sparse-to-dense alignment* strategy enabling both consistent global structure with implicit loop closures and highly-accurate fine-scale pose alignment to facilitate local surface detail.

(3) A new RGB-D re-integration strategy to enable on-the-fly and continuous 3D model updates when refined global pose estimates are available.

(4) Large-scale reconstruction of geometry and texture, demonstrating model refinement in revisited areas, recovery from tracking failures, and robustness to drift and continuous loop closures.

## 2 RELATED WORK

There has been extensive work on 3D reconstruction over the past decades. Key to high-quality 3D reconstruction is the choice of underlying representation for fusing multiple sensor measurements. Approaches range from unstructured point-based representations [18, 22, 41, 50, 54], 2.5D depth map [30, 32] or height-field [14] methods, to volumetric approaches, based on occupancy grids [7, 56] or implicit surfaces [5, 19]. While each has trade-offs, volumetric methods based on implicit truncated signed distance fields (TSDFs) have become the de facto method for highest quality reconstructions; e.g., [11, 13, 26]. They model continuous surfaces, systematically regularize noise, remove the need for explicit topology bookkeeping, and efficiently perform incremental updates. The most prominent recent example is KinectFusion [20, 34] where real-time volumetric fusion of smaller scenes was demonstrated.

One inherent issue with these implicit volumetric methods is their lack of scalability due to reliance on a uniform grid. This has become a focus of much recent research [3, 22, 37, 39, 40, 45, 51, 58, 59], where real-time efficient data structures for volumetric fusion have been proposed. These exploit the sparsity in the TSDF representation to create more efficient spatial subdivision strategies. While this allows for volumetric fusion at scale, pose estimates suffer from drift, causing distortions in the 3D model. Even small pose errors, seemingly negligible on a small local scale, can accumulate to dramatic error in the final 3D model [37]. Zhang et al. [59] use planar structural priors and repeated object detection to reduce the effect of drift; however, they do not detect loop closures or use color data, which makes tracking difficult in open or planar areas, or very cluttered scenes.

Most of the research on achieving globally consistent 3D models at scale from RGB-D input requires offline processing and access to all input frames. [4, 27, 60–62] provide for globally consistent models by optimizing across the entire pose trajectory, but require minutes or even hours of processing time, meaning *real-time* revisiting or refinement of reconstructed areas is infeasible.

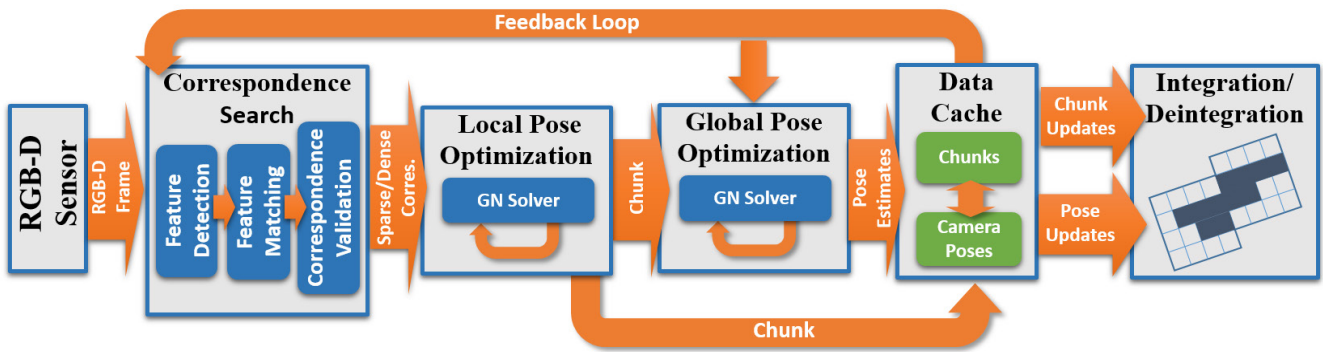

Fig. 2. Our global pose optimization takes as input the RGB-D stream of a commodity sensor, detects pairwise correspondences between the input frames, and performs a combination of local and global alignment steps using sparse and dense correspondences to compute per-frame pose estimates.

Real-time, drift-free pose estimation is a key focus in the simultaneous localization and mapping (SLAM) literature. Many real-time monocular RGB methods have been proposed, including sparse methods [24], semi-dense [10, 12] or direct methods [9, 31]. Typically these approaches rely on either pose-graph optimization [25] or bundle adjustment [48], minimizing reprojection error across frames and/or distributing the error across the graph. While impressive tracking results have been shown using only monocular RGB sensors, these approaches do not generate detailed dense 3D models, which is the aim of our work.

MonoFusion [38] augments sparse SLAM bundle adjustment with dense volumetric fusion, showing compelling monocular results but on small-scale scenes. Real-time SLAM approaches typically first estimate poses *frame-to-frame* and perform correction in a background thread (running slower than real-time rates; e.g., 1Hz). In contrast, DTAM [35] uses the concept of *frame-to-model* tracking (from KinectFusion [20, 34]) to estimate the pose directly from the reconstructed dense 3D model. This omits the need for a correction step, but clearly does not scale to larger scenes.

Pose estimation from range data typically is based on variants of the iterative closest point (ICP) algorithm [2, 42]. In practice, this makes tracking extremely brittle and has led researchers to explore either the use of RGB data to improve frame-to-frame tracking [52] or the use of global pose estimation correction, including pose graph optimization [44], loop closure detection [53], incremental bundle adjustement [11, 54], or recovery from tracking failures by image or keypoint-based relocalization [15, 49].

These systems are state-of-the-art in terms of online correction of both pose and underlying 3D model. However, they either require many seconds or even minutes to perform online optimization [11, 53]; assume very specific camera trajectories to detect explicit loop closures limiting free-form camera motions and scanning [53]; rely on computing optimized camera poses prior to fusion limiting the ability to refine the model afterwards [44], or use point-based representations that limit quality and lack general applicability where continuous surfaces are needed [54].

## 3 METHOD OVERVIEW

The core of our approach is an efficient global pose optimization algorithm which operates in unison with a large-scale, real-time 3D reconstruction framework; see Fig. 2. At every frame, we continuously run pose optimization and update the reconstruction according to the newly-computed pose estimates. We do not strictly rely on temporal coherence, allowing for free-form camera paths, instantaneous relocalization, and frequent revisiting of the same scene region. This makes our approach robust towards sensor occlusion, fast frame-to-frame motions and featureless regions.

We take as input the RGB-D stream captured by a commodity depth sensor. To obtain global alignment, we perform a sparse-then-dense global pose optimization: we use a set of sparse feature correspondences to obtain a coarse global alignment, as sparse features inherently provide for loop closure detection and relocalization. This alignment is then refined by optimizing for dense photometric and geometric consistency. Sparse correspondences are established through pairwise Scale-Invariant Feature Transform (SIFT) [28] feature correspondences between all input frames (see Sec. 4.1). That is, detected SIFT keypoints are matched against all previous frames, and carefully filtered to remove mismatches, thus avoiding false loop closures (see Sec. 4.1.1).

To make real-time global pose alignment tractable, we perform a hierarchical local-to-global pose optimization (see Sec. 4.2) using the filtered frame correspondences. On the first hierarchy level, every consecutive *n* frames compose a *chunk*, which is locally pose optimized under the consideration of its contained frames. On the second hierarchy level, all chunks are correlated with respect to each other and globally optimized. This is akin to hierarchical submapping [29]; however, instead of analyzing global connectivity once all frames are available, our new method forms *chunks* based on the current temporal window. Note that this is our only temporal assumption; between chunks there is no temporal reliance.

This hierarchical two-stage optimization strategy reduces the number of unknowns per optimization step and ensures our method scales to large scenes. Pose alignment on both levels is formulated as energy minimization problem in which both the filtered sparse correspondences, as well as dense photometric and geometric constraints are considered (see Sec. 4.3). To solve this highly-nonlinear optimization problem on both hierarchy levels, we employ a fast data-parallel GPU-solver tailored to the problem (see Sec. 4.4).

A dense scene reconstruction is obtained using a sparse volumetric representation and fusion [37], which scales to large scenes in real-time. The continuous change in the optimized global poses necessitates continuous updates to the global 3D scene representation (see Sec. 5). A key novelty is to allow for symmetric on-the-fly reintegration of RGB-D frames. In order to update the pose of a frame with an improved estimate, we remove the RGB-D image at the old pose with a new real-time *de-integration* step, and *reintegrate* it at the new pose. Thus, the volumetric model continuously improves as more RGB-D frames and refined pose estimates become available; e.g., if a loop is closed (cf. Fig. 13).

# 4 GLOBAL POSE ALIGNMENT

We first describe the details of our real-time global pose optimization strategy, which is the foundation for *online*, globally-consistent 3D reconstruction. Input to our approach is the live RGB-D stream $S = \{f_i = (C_i, \mathcal{D}_i)\}_i$ captured by a commodity sensor. We assume spatially and temporally aligned color $C_i$ and depth data $\mathcal{D}_i$ at each frame, captured at 30Hz and $640 \times 480$ pixel resolution. The goal is to find a set of 3D correspondences between the frames in the input sequence, and then find an optimal set of rigid camera transforms $\{\mathcal{T}_i\}$ such that all frames align as best as possible. The transformation $\mathcal{T}_i(\mathbf{p}) = \mathbf{R}_i\mathbf{p} + \mathbf{t}_i$ (rotation $\mathbf{R}_i$, translation $\mathbf{t}_i$) maps from the local camera coordinates of the $i$-th frame to the world space coordinate system; we assume the first frame defines the world coordinate system.

## 4.1 Feature Correspondence Search

In our framework, we first search for sparse correspondences between frames using efficient feature detection, feature matching, and correspondence filtering steps. These sparse correspondences are later used in tandem with dense photometric correspondences, but since accurate sparse correspondences are crucial to attaining the basin of convergence of the dense optimization, we elaborate on their search and filtering below. For each new frame, SIFT features are detected and matched to the features of all previously seen frames. We use SIFT as it accounts for the major variation encountered during hand-held RGB-D scanning, namely: image translation, scaling, and rotation. Potential matches between each pair of frames are then filtered to remove false positives and produce a list of valid pairwise correspondences as input to global pose optimization. Our correspondence search is performed entirely on the GPU, avoiding the overhead of copying data (e.g., feature locations, descriptors, matches) to the host. We compute SIFT keypoints and descriptors at $4 - 5$ ms per frame, and match a pair of frames in $\approx 0.05$ms (in parallel). We can thus find full correspondences in real-time against up to over 20K frames, matched in a hierarchical fashion, for every new input RGB-D image.

### 4.1.1 Correspondence Filtering.
To minimize outliers, we filter the sets of detected pairwise correspondences based on geometric and photometric consistency. Note that further robustness checks are built into the optimization (not described in this section; see Sec. 4.4.1 for details).

**Key Point Correspondence Filter** For a pair of frames $f_i$ and $f_j$ with detected corresponding 3D points $P$ from $f_i$, and $Q$ from $f_j$, the key point correspondence filter finds a set of correspondences which exhibit a stable distribution and a consistent rigid transform. Correspondences are greedily aggregated (in order of match distance); for each newly added correspondence, we compute the rigid transform $\mathcal{T}_{ij}(\mathbf{p}) = (\mathcal{T}_j^{-1} \circ \mathcal{T}_i)(\mathbf{p})$, which minimizes the RMSD between the current set of correspondences $P_{cur}$ and $Q_{cur}$, using the Kabsch algorithm [16, 21]. We further check whether this is an ambiguously determined transform (e.g, the correspondences lie on a line or exhibit rotational symmetry) by performing a condition analysis of the covariance of points of $P_{cur}$ and $Q_{cur}$ as well as the cross-covariance between $P_{cur}$ and $Q_{cur}$; if any of these condition numbers are high ($> 100$) the system is considered unstable. Thus, if the re-projection error under $\mathcal{T}_{ij}$ is high (max residual $> 0.02$m) or the condition analysis determines instability, then correspondences are removed (in order of re-projection error) until this is not the case anymore or there are too few correspondences to determine a rigid transform. If the resulting set of correspondences for $f_i$ and $f_j$ do not produce a valid transform, all correspondences between $f_i$ and $f_j$ are discarded.

**Surface Area Filter** In addition, we check that the surface spanned by the features is large enough, as correspondences spanning small physical size are prone to ambiguity. For frames $f_i$ and $f_j$, we estimate the surface areas spanned by the 3D keypoints $P$ of $f_i$ and by the 3D keypoints $Q$ of $f_j$. For each set of 3D points, we project them into the plane given by their two principal axes, with surface area given by the 2D oriented bounding box of the resulting projected points. If the areas spanned by $P$ and $Q$ are insufficient ($< 0.032 \text{m}^2$), the set of matches is deemed ambiguous and discarded.

**Dense Verification** Finally, we perform a dense two-sided geometric and photometric verification step. For frames $f_i$ and $f_j$, we use the computed relative transform $\mathcal{T}_{ij}$ from the key point correspondence filter to align the coordinate systems of $f_i$ and $f_j$. We measure the average depth discrepancy, normal deviation and photoconsistency of the re-projection in both directions in order to find valid pixel correspondences, and compute the re-projection error of these correspondences. For efficiency reasons, this step is performed on filtered and downsampled input frames of size $w' \times h' = 80 \times 60$. Note that when a new RGB-D image $f_i$ arrives, its filtered and downsampled color intensity $C_i^{low}$ and depth $\mathcal{D}_i^{low}$ are cached for efficiency. The camera space positions $P_i^{low}$ and normals $N_i^{low}$ of each $\mathcal{D}_i^{low}$ are also computed and cached per frame. With $\pi$ denoting the camera intrinsics for the downsampled images, the total re-projection error from $f_i$ to $f_j$ is:

$$E_r(f_i, f_j) = \sum_{x,y} \left\| \mathcal{T}_{ij}(\mathbf{p}_{i,x,y}) - \mathbf{q}_{j,x,y} \right\|_2.$$

Here, $\mathbf{p}_{i,x,y} = P_i^{low}(x,y)$ and $\mathbf{q}_{j,x,y} = P_j^{low}(\pi^{-1}(\mathcal{T}_{ij}\mathbf{p}_{i,x,y}))$. However, this is sensitive to occlusion error, so we discard correspondences with high depth discrepancy, normal deviation, or lack of photoconsistency. That is, the potential correspondence at pixel location $(x, y)$ is considered valid if the following conditions hold:

$$\left\| \mathcal{T}_{ij}(\mathbf{p}_{i,x,y}) - \mathbf{q}_{j,x,y} \right\|_2 < \tau_d$$
$$(\mathcal{T}_{ij}(\mathbf{n}_{i,x,y})) \cdot \mathbf{n}_{j,x,y} > \tau_n$$
$$\left\| C_i^{low}(x,y) - C_j^{low}(x,y) \right\|_1 < \tau_c$$

Matches between $f_i$ and $f_j$ are invalidated in the case of excessive re-projection error ($> 0.075$m) or insufficient valid correspondences ($< 0.02w'h'$), and we use $\tau_d = 0.15$m, $\tau_n = 0.9$, $\tau_c = 0.1$. This check is efficiently implemented with a single kernel call, such that each thread block handles one pair of images, with re-projection computed through local reductions.

If all checks are passed, the correspondences are added to the valid set, which is used later on for pose optimization. We only consider a frame-to-frame match if the valid set comprises at least $N_{min}$ correspondences. Note that $N_{min} = 3$ is sufficient to define a valid frame-to-frame transform; however, we found $N_{min} = 5$ to be a good compromise between precision and recall.

### 4.2 Hierarchical Optimization

In order to run at real-time rates on up to tens of thousands of RGB-D input frames, we apply a hierarchical optimization strategy. The input sequence is split into short *chunks* of consecutive frames. On the lowest hierarchy level, we optimize for local alignments within a *chunk*. On the second hierarchy level, chunks are globally aligned against each other, using representative *keyframes* with associated features per chunk.

**Local Intra-Chunk Pose Optimization** Intra-chunk alignment is based on chunks of $N_{chunk} = 11$ consecutive frames in the input RGB-D stream; adjacent chunks overlap by 1 frame. The goal of local pose optimization is to compute the best intra-chunk alignments $\{\mathcal{T}_i\}$, relative to the first frame of the chunk, which locally defines the reference frame. To this end, valid feature correspondences are searched between all pairs of frames of the chunk, and then the energy minimization approach described in Sec. 4.3 is applied, jointly considering both these feature correspondences *and* dense photometric and geometric matching. Since each chunk only contains a small number of consecutive frames, the pose variation within the chunk is small, and we can initialize each of the $\mathcal{T}_i$ to the identity matrix. To ensure that the local pose optimization result after convergence is sufficiently accurate, we apply the Dense Verification test (see Sec. 4.1.1) to each pair of images within the chunk using the optimized local trajectory. If the re-projection error is too large for any pair of images ($> 0.05$m), the chunk is discarded and not used in the global optimization.

**Per-Chunk Keyframes** Once a chunk has been completely processed, we define the RGB-D data from the first frame in the chunk to be the chunk's *keyframe*. We also compute a representative aggregate *keyframe feature set*. Based on the optimized pose trajectory of the chunk, we compute a coherent set of 3D positions of the intra-chunk feature points in world space. These 3D positions may contain multiple instances of the same real-world point, found in separate pairwise frame matches. Thus, to obtain the keyframe feature set, we aggregate the feature point instances that have previously found (intra-chunk) matches. Those that coincide in 3D world space ($< 0.03$m) are merged to one best 3D representative in the least squares sense. This keyframe feature set is projected into the space of the keyframe using the transformations from the frames of origin, resulting in a consistent set of feature locations and depths. Note that once this global keyframe and keyframe feature set is created, the chunk data (i.e., intra-chunk features, descriptors, correspondences) can be discarded as it is not needed in the second layer pose alignment.

**Global Inter-Chunk Pose Optimization** Sparse correspondence search and filtering between global keyframes is analogous to that within a chunk, but on the level of all keyframes and their feature sets. If a global keyframe does not find any matches to previously seen keyframes, it is marked as invalid but kept as a candidate, allowing for re-validation when it finds a match to a keyframe observed in the future. The global pose optimization computes the best global alignments $\{\mathcal{T}_i\}$ for the set of all global keyframes, thus aligning all chunks globally. Again, the same energy minimization approach from Sec. 4.3 is applied using both sparse and dense constraints. Intra-chunk alignment runs after each new global keyframe has found correspondences. The pose for a global keyframe is initialized with the delta transform computed by the corresponding intra-chunk optimization, composed with the previous global keyframe pose. After the intra-chunk transforms have been computed, we obtain globally consistent transforms among all input frames by applying the corresponding delta transformations (from the local optimization) to all frames in a chunk.

### 4.3 Pose Alignment as Energy Optimization

Given a set of 3D correspondences between a set of frames $\mathbf{S}$ (frames in a chunk or keyframes, depending on hierarchy level), the goal of pose alignment is to find an optimal set of rigid camera transforms $\{\mathcal{T}_i\}$ per frame $i$ (for simpler notation, we henceforth write $i$ for $f_i$) such that all frames align as best as possible. We parameterize the $4 \times 4$ rigid transform $\mathcal{T}_i$ using matrix exponentials based on skew-symmetric matrix generators [33], which yields fast convergence. This leaves 3 unknown parameters for rotation, and 3 for translation. For ease of notation, we stack the degrees of freedom for all $|\mathbf{S}|$ frames in a parameter vector:

$$\mathcal{X} = (\mathbf{R}_0, \mathbf{t}_0, \dots, \mathbf{R}_{|\mathbf{S}|}, \mathbf{t}_{|\mathbf{S}|})^T = (x_0, \dots, x_N)^T .$$

Here, $N$ is the total number of variables $x_i$. Given this notation, we phrase the alignment problem as a variational non-linear least squares minimization problem in the unknown parameters $\mathcal{X}$. To this end, we define the following alignment objective, which is based on sparse features *and* dense photometric and geometric constraints:

$$E_{\text{align}}(\mathcal{X}) = w_{\text{sparse}}E_{\text{sparse}}(\mathcal{X}) + w_{\text{dense}}E_{\text{dense}}(\mathcal{X}).$$

Here, $w_{\text{sparse}}$ and $w_{\text{dense}}$ are weights for the sparse and dense matching terms, respectively. $w_{\text{dense}}$ is linearly increased; this allows the sparse term to first find a good global structure, which is then refined with the dense term (as the poses fall into the basin of convergence of the dense term, it becomes more reliable), thus achieving coarse-to-fine alignment. Note that depending on the optimization hierarchy level, the reference frame is the first frame in the chunk (for intra-chunk alignment), or the first frame in the entire input sequence (for global inter-chunk alignment). Hence, the reference transform $\mathcal{T}_0$ is not a free variable and left out from the optimization.

**Sparse Matching** In the sparse matching term, we minimize the sum of distances between the world space positions over all feature correspondences between all pairs of frames in $\mathbf{S}$:

$$E_{\text{sparse}}(\mathcal{X}) = \sum_{i=1}^{|\mathbf{S}|} \sum_{j=1}^{|\mathbf{S}|} \sum_{(k,l) \in \mathbf{C}(i,j)} \left\| \mathcal{T}_i \mathbf{p}_{i,k} - \mathcal{T}_j \mathbf{p}_{j,l} \right\|_2^2 .$$

Here, $\mathbf{p}_{i,k}$ is the $k$-th detected feature point in the $i$-th frame. $\mathbf{C}_{i,j}$ is the set of all pairwise correspondences between the $i$-th and the $j$-th frame. Geometrically speaking, we seek the best rigid transformations $\mathcal{T}_i$ such that the Euclidean distance over all the detected feature matches is minimized.

**Dense Matching** We additionally use dense photometric and geometric constraints for fine-scale alignment. To this end, we exploit the dense pixel information of each input frame's color $\mathcal{C}_i$ and depth $\mathcal{D}_i$. Evaluating the dense alignment is computationally more expensive than the previous sparse term. We therefore evaluate it on a restricted set $\mathbf{E}$ of frame pairs, $\mathbf{E}$ contains a frame pair $(i, j)$ if their camera angles are similar (within $60°$, to avoid glancing angles of the same view) and they have non-zero overlap with each other; this can be thought of as encoding the edges $(i, j)$ of a sparse matching graph. The optimization for both dense photometric and geometric alignment is based on the following energy:

$$E_{\text{dense}}(\mathcal{T}) = w_{\text{photo}} E_{\text{photo}}(\mathcal{T}) + w_{\text{geo}} E_{\text{geo}}(\mathcal{T}).$$

Here, $w_{\text{photo}}$ is the weight of the photometric term and $w_{\text{geo}}$ of the geometric term, respectively. For the dense photo-consistency term, we evaluate the error on the gradient $\mathcal{I}_i$ of the luminance of $\mathcal{C}_i$ to gain robustness against lighting changes:

$$E_{\text{photo}}(\mathcal{X}) = \sum_{(i,j) \in \mathbf{E}} \sum_{k=0}^{|\mathcal{I}_i|} \left\| \mathcal{I}_i(\pi(\mathbf{d}_{i,k})) - \mathcal{I}_j(\pi(\mathcal{T}_j^{-1} \mathcal{T}_i \mathbf{d}_{i,k})) \right\|_2^2.$$

Here, $\pi$ denotes the perspective projection, and $\mathbf{d}_{i,k}$ is the 3D position associated with the $k$-th pixel of the $i$-th depth frame. Our geometric alignment term evaluates a point-to-plane metric to allow for fine-scale alignment in the tangent plane of the captured geometry:

$$E_{\text{geo}}(\mathcal{X}) =$$
$$\sum_{(i,j) \in \mathbf{E}} \sum_{k=0}^{|\mathcal{D}_i|} \left[ \mathbf{n}_{i,k}^T \left( \mathbf{d}_{i,k} - \mathcal{T}_i^{-1} \mathcal{T}_j \pi^{-1} \left( \mathcal{D}_j \left( \pi(\mathcal{T}_j^{-1} \mathcal{T}_i \mathbf{d}_{i,k}) \right) \right) \right) \right]^2.$$

Here, $\mathbf{n}_{i,k}$ is the normal of the $k$-th pixel in the $i$-th input frame. Correspondences that project outside of the input frame are ignored, and we apply ICP-like pruning based on distance and normal constraints after each optimization step. For the dense photometric and geometric constraints, we downsample $\mathcal{I}_i$ and $\mathcal{D}_i$, to $80 \times 60$ pixels (using the same cached frames as for the dense verification filter). Note that for the global pose optimization, the result of optimizing densely at every keyframe is effectively reset by the sparse correspondence optimization, since the 3D positions of the correspondences are fixed. Thus we only perform the dense global keyframe optimization after the user has indicated the end of scanning.

## 4.4 Fast and Robust Optimization Strategy

The described global pose alignment objective is a non-linear least squares problem in the unknown extrinsic camera parameters. Since our goal is *online* and *global* camera pose optimization for long scanning sequences with over twenty thousand frames, an efficient, yet effective, optimization strategy is required. To face this challenge, we implement a data-parallel GPU-based non-linear iterative solver similar to the work of Zollhöfer et al. [64]. However, the unique sparsity pattern associated with the global alignment objective requires a different parallelization strategy and prohibits the use of previous GPU-based solvers [55, 63, 64]. Our approach is based on the Gauss-Newton method, which only requires first order derivatives and exhibits quadratic convergence close to the optimum, which is beneficial due to our incremental optimization scheme. We find the best pose parameters $\mathcal{X}^*$ by minimizing the proposed highly non-linear least squares objective using this method:

$$\mathcal{X}^* = \underset{\mathcal{X}}{\operatorname{argmin}} \, E_{align}(\mathcal{X}) .$$

For ease of notation, we reformulate the objective in the following canonical least-squares form:

$$E_{align}(\mathcal{X}) = \sum_{i=1}^{R} r_i(\mathcal{X})^2 .$$

This is done by re-naming the $R = 3N_{corr} + |\mathbf{E}| \cdot (|\mathcal{I}_i| + |\mathcal{D}_i|)$ terms of the energy appropriately. Here, $N_{corr}$ is either the total number of inter-chunk sparse correspondences for inter-chunk alignment, or per-chunk sparse correspondences for intra-chunk alignment. The notation can be further simplified by defining a vector field $\mathbf{F} : \mathbb{R}^N \to \mathbb{R}^R$ that stacks all scalar residuals:

$$\mathbf{F}(\mathcal{X}) = [ \, \dots, r_i(\mathcal{X}), \, \dots \, ]^T .$$

With this notation, $E_{refine}$ can be expressed in terms of the squared Euclidean length of $\mathbf{F}(\mathcal{X})$:

$$E_{refine}(\mathcal{X}) = ||\mathbf{F}(\mathcal{X})||_2^2.$$

Gauss-Newton is applied via a local linear approximation of $\mathbf{F}$ at the last solution $\mathcal{X}^{k-1}$ using first-order Taylor expansion:

$$\mathbf{F}(\mathcal{X}^k) = \mathbf{F}(\mathcal{X}^{k-1}) + \mathbf{J_F}(\mathcal{X}^{k-1}) \cdot \Delta\mathcal{X}, \ \Delta\mathcal{X} = \mathcal{X}^k - \mathcal{X}^{k-1} .$$

Here, $\mathbf{J_F}$ denotes the Jacobian of $\mathbf{F}$. By substituting $\mathbf{F}$ with this local approximation, the optimal parameter update $\Delta\mathcal{X}$ is found by solving a linear least squares problem:

$$\Delta\mathcal{X}^* = \underset{\Delta\mathcal{X}}{\operatorname{argmin}} \underbrace{||\mathbf{F}(\mathcal{X}^{k-1}) + \mathbf{J_F}(\mathcal{X}^{k-1}) \cdot \Delta\mathcal{X}||_2^2}_{E_{lin}(\Delta\mathcal{X})} .$$

To obtain the minimizer $\Delta\mathcal{X}^*$, we set the corresponding partial derivatives $\frac{dE_{lin}}{d\Delta\mathcal{X}_i}(\Delta\mathcal{X}_i^*) = 0$, $\forall i$ to zero, which yields the following system of linear equations:

$$\mathbf{J_F}(\mathcal{X}^{k-1})^T \mathbf{J_F}(\mathcal{X}^{k-1}) \cdot \Delta\mathcal{X}^* = -\mathbf{J_F}(\mathcal{X}^{k-1})^T \mathbf{F}(\mathcal{X}^{k-1}) .$$

To solve the system, we use a GPU-based data-parallel Preconditioned Conjugate Gradient (PCG) solver with Jacobi preconditioner. Based on the iterative solution strategy, the sparsity of the system matrix $\mathbf{J_F}(\mathcal{X}^{k-1})^T \mathbf{J_F}(\mathcal{X}^{k-1})$ can be exploited. For the sparse term, we never explicitly compute this matrix, but compute the non-zero entries, if required, on-the-fly during the PCG iterations.

Gauss-Newton iterates this process of locally linearizing the energy and solving the associated linear least squares problem starting from an initial estimate $\mathcal{X}_0$ until convergence. We warm-start the optimization based on the result obtained in the last frame.

In contrast to Zollhöfer [64], instead of using a reduction based on two kernels to compute the optimal step size and update for the descent direction, we use a single kernel running a combination of warp scans based on the *shuffle* intrinsic and global memory atomics to accumulate the final result. This turned out to be much faster for our problem size.

The central operation of the PCG algorithm is the multiplication of the system matrix with the current descent direction.

Let us first consider the sparse feature term. To avoid fill-in, we multiply the system matrix incrementally based on two separate kernel calls: the first kernel multiplies $\mathbf{J_F}$ and the computed intermediate result is then multiplied by $\mathbf{J_F^T}$ in the second kernel call. For example, at the end of the *apt0* sequence (see Fig. 3, bottom), $\mathbf{J_F}$ has about 105K rows (residuals) and 5K columns (unknowns). In all operations, we exploit the sparse structure of the matrix, only performing operations which will lead to a non-zero result. Since $\mathbf{J_F}$ and $\mathbf{J_F^T}$ have very different row-wise sparsity patterns, using two different kernel calls helps to fine tune the parallelization approach to the specific requirements.

More specifically, for the sparse term, each row of $\mathbf{J_F}$ encodes exactly one pairwise correspondence, depending on at most 2 extrinsic camera poses or $2 \times 6 = 12$ non-zero matrix entries. Due to the low number of required operations, the matrix-vector product can be readily computed by assigning one dedicated thread to each 3D block row; i.e., handling the $x$-, $y$-, and $z$-residuals of one correspondence. This is beneficial, since the different dimensions share common operations in the evaluation of $\mathbf{F}$ and $\mathbf{J_F}$. In contrast, $\mathbf{J}^T$ has exactly one row per unknown. The number of non-zero entries in each row is equivalent to the number of correspondences involving the frame associated with the unknown. For longer scanning sequences, this can easily lead to several thousand entries per row. To reduce the amount of memory reads and compute of each thread, we opted for a reduction-based approach to compute the matrix-vector products. We use one block of size $N_{block} = 256$ to compute each row-wise dot product. Each warp of a block performs a warp-reduction based on the *shuffle* intrinsic and the final per-warp results are combined based on shared memory atomics. For computing the multiplication with $\mathbf{J_F^T}$, we pre-compute auxiliary lists that allow lookup to all correspondences that influence a certain variable. This table is filled based on a kernel that has one thread per correspondence and adds entries to the lists corresponding to the involved variables. The per-list memory is managed using *atomic* counters. We recompute this table if the set of active correspondences changes.

For the dense photometric and geometric alignment terms, the number of associated residuals is considerably higher. Since the system matrix is fixed during the PCG steps, we pre-compute it at the beginning of each non-linear iteration. The required memory is preallocated and we update only the non-zero entries via scattered writes. Note that we only require a few writes, since we perform the local reductions in shared memory.

### 4.4.1 Correspondence and Frame Filtering.
As an additional safeguard to make the optimization robust against potential correspondence outliers, which were mistakenly considered to be valid, we perform correspondence and frame filtering after each optimization finishes. That is, we determine the maximum residual $r_{max} = \max_i r_i(\mathcal{X})$ using a parallel reduction on the GPU, with the final max computation performed on the CPU. If $r_{max} > 0.05m$, we remove all correspondences between the two frames $i$ and $j$ associated with the correspondence which induces $r_{max}$. Note that all correspondences between $i$ and $j$ are removed, in order to minimize the number of times the optimization has to run in order to prune all bad correspondences. Additionally, if a frame has no correspondence to any other frame, it is implicitly removed from the optimization and marked as invalid.

Note that the vast majority of false loop closures are filtered out through the verification steps (Sec. 4.1.1), and the optimization pruning effectively removes the rest. Table 2 provides a detailed overview of the effects of these filtering steps.

## 5 DYNAMIC 3D RECONSTRUCTION

Key to live, globally consistent reconstruction is updating the 3D model based on newly-optimized camera poses. We thus monitor the continuous change in the poses of each frame to update the volumetric scene representation through *integration* and *de-integration* of frames. Based on this strategy, errors in the volumetric representation due to accumulated drift or dead reckoning in feature-less regions can be fixed as soon as better pose estimates are available.

### 5.1 Scene Representation

Scene geometry is reconstructed by incrementally fusing all input RGB-D data into an implicit truncated signed distance (TSDF) representation, following Curless and Levoy [5]. The TSDF is defined over a volumetric grid of voxels; to store and process this data, we employ the state-of-the-art sparse volumetric voxel hashing approach proposed by Nießner et al. [37]. This approach scales well to the scenario of large-scale surface reconstruction, since empty space neither needs to be represented nor addressed; the TSDF is stored in a sparse volumetric grid based on spatial hashing. Following the original approach, we also use voxel blocks of $8 \times 8 \times 8$ voxels. In contrast to the work of Nießner et al. [37], we allow for RGB-D frames to be both *integrated* into the TSDF as well as *de-integrated* (i.e., adding and removing frames from the reconstruction). In order to allow for pose updates, we also ensure that these two operations are symmetric; i.e., one inverts the other.

### 5.2 Integration and De-integration

Integration of a depth frame $\mathcal{D}_i$ occurs as follows. For each voxel, $\mathbf{D(v)}$ denotes the signed distance of the voxel, $\mathbf{W(v)}$ the voxel weight, $d_i(\mathbf{v})$ the projective distance (along the $z$ axis) between a voxel and $\mathcal{D}_i$, and $w_i(\mathbf{v})$ the integration weight for a sample of $\mathcal{D}_i$. For data integration, each voxel is then updated by

$$\mathbf{D'(v)} = \frac{\mathbf{D(v)W(v)} + w_i(\mathbf{v})d_i(\mathbf{v})}{\mathbf{W(v)} + w_i(\mathbf{v})}, \quad \mathbf{W'(v)} = \mathbf{W(v)} + w_i(\mathbf{v}).$$

We can reverse this operation to de-integrate a frame. Each voxel is then updated by

$$D'(v) = \frac{D(v)W(v) - w_i(v)d_i(v)}{W(v) - w_i(v)}, \quad W'(v) = W(v) - w_i(v).$$

We can thus update a frame in the reconstruction by de-integrating it from its original pose and integrating it with a new pose. This is crucial for obtaining high-quality reconstructions in the presence of loop closures and revisiting, since the already integrated surface measurements must be adapted to the continuously changing stream of pose estimates.

### 5.3 Managing Reconstruction Updates

Each input frame is stored with its associated depth and color data, along with two poses: its *integrated* pose, and its *optimized* pose. The integrated pose is the one used currently in the reconstruction, and is set whenever a frame gets integrated. The optimized pose stores the (continually-changing) result of the pose optimization.

When an input frame arrives, we aim to integrate it into the reconstruction as quickly as possible, to give the user or robot instantaneous feedback of the 3D model. Since the global optimization is not run for each frame but for each chunk, an optimized pose may not immediately be available and we must obtain an initial transform by other means. We compute this initial transform by composing the frame-to-frame transform from the key point correspondence filter with the newest available optimized transform.

In order to update the reconstruction with the most pertinent optimization updates, we sort the frames in descending order by the difference between the integrated transform and the optimized transform. The integrated transform and optimized transform are parameterized by 6 DOFs: $\alpha, \beta, \gamma$ (here, we use Euler angles in radians) describing the rotation, and $x, y, z$ (in meters) describing the translation. Then the distance between the integrated transform $t_{int} = (\alpha_i, \beta_i, \gamma_i, x_i, y_i, z_i)$ and the optimized transform $t_{opt} = (\alpha_o, \beta_o, \gamma_o, x_o, y_o, z_o)$ is defined to be $\|s * t_{int} - s * t_{opt}\|_2$ where $s = (2, 2, 2, 1, 1, 1)$ is multiplied element-wise to bring the rotations and translations closer in scale. For each new input frame, we de-integrate and integrate the $N_{fix} = 10$ frames from the top of the list. This allows us to dynamically update the reconstruction to produce a globally-consistent 3D reconstruction.

## 6 RESULTS

For live scanning, we use a *Structure Sensor*[2] mounted to an iPad Air. The RGB-D stream is captured at 30Hz with a color and depth resolution of $640 \times 480$. Note that we are agnostic to the type of used depth sensor. We stream the captured RGB-D data via a wireless network connection to a desktop machine that runs our global pose optimization and reconstructs a 3D model in real-time. Visual feedback of the reconstruction is streamed live to the iPad to aid in the scanning process. To reduce the required bandwidth, we use data compression based on *zlib* for depth and *jpeg* compression for color. We implemented our global pose alignment framework using the CUDA 7.0 architecture. Reconstruction results of scenes captured using our live system are shown in Fig. 1 and 3 as well as in the supplementary video. The completeness of the various

---

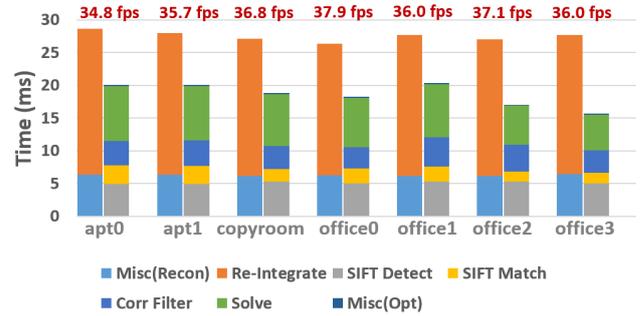

Fig. 4. Performance Evaluation: our proposed pipeline runs at well beyond 30Hz for all used test sequences. The computations are split up over two GPUs (left bar Titan X, right bar Titan Black).

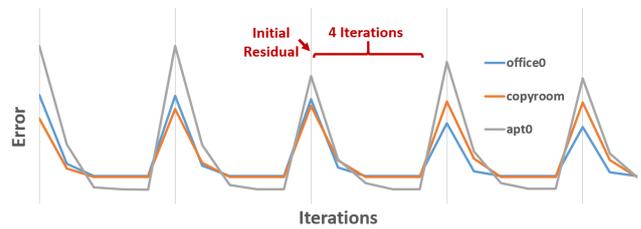

Fig. 5. Convergence analysis of the global keyframe optimization (log scale): peaks correspond to new global keyframes. Only a few iterations are required for convergence.

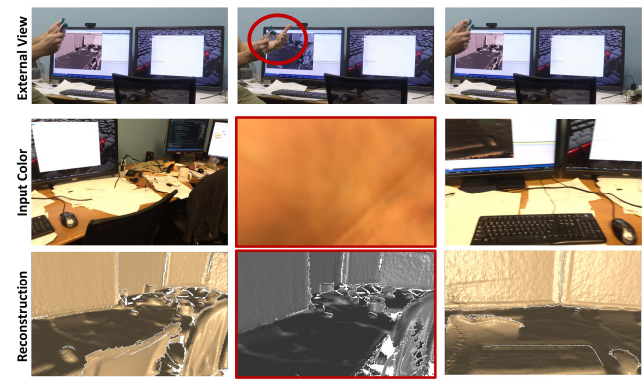

Fig. 6. Recovery from tracking failure: our method is able to detect (gray overlay) and recover from tracking failure; i.e., if the sensor is occluded or observes a featureless region.

large-scale indoor scenes (4 offices, 2 apartments, 1 copyroom, with up to 95m camera trajectories), their alignment without noticeable camera drift, and the high local quality of geometry and texture are on par with even *offline* approaches. This also demonstrates that our global pose alignment strategy scales well to large spatial extents and long sequences (over 20,000 frames).

*Qualitative Comparison.* First, we compare to the online 3D reconstruction approach of Nießner et al. [37], see Fig. 12. In contrast

---

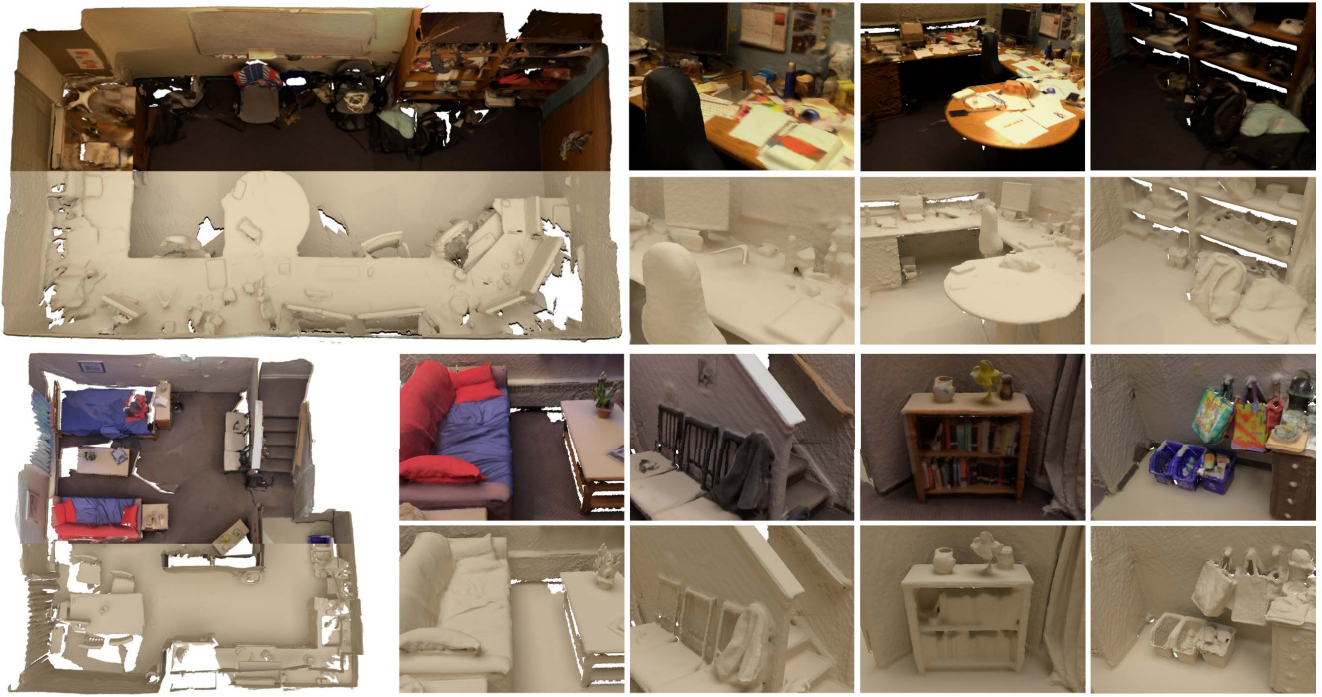

Fig. 3. Large-scale reconstruction results: our proposed real-time global pose optimization outperforms current state-of-the-art online reconstruction systems. The globally aligned 3D reconstructions are at a quality that was previously only attainable offline. Note the completeness of the scans, the global alignment without noticeable camera drift and the high local quality of the reconstructions in both geometry and texture. Scans comprise thousands of input frames, include revisiting and many loop closures.

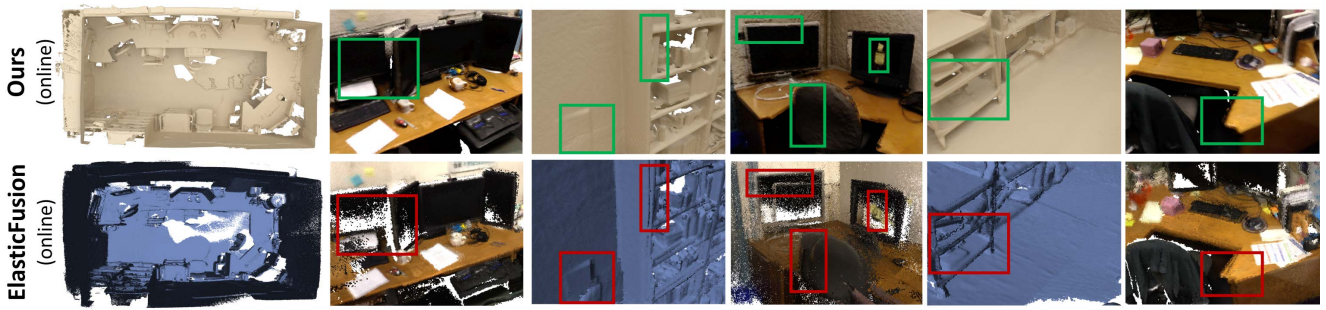

Fig. 7. Our proposed real-time global pose optimization (top) outperforms the method of Whelan et al. [54] (bottom) in terms of scan completeness and alignment accuracy. Note, we generate a high-quality surface mesh, while the competing approach only outputs a pointcloud.

to their work, which builds on frame-to-model tracking and suffers from the accumulation of camera drift, we are able to produce drift-free reconstructions at high fidelity. Our novel global pose optimization framework implicitly handles loop closure, recovers from tracking failures, and reduces geometric drift. Note that most real-time fusion methods (e.g., [3, 20, 34, 37]) share the same frame-to-model ICP tracking algorithm, and therefore suffer from notable drift. Fig. 7 and 9 show a comparison of our approach with the online *ElasticFusion* approach of Whelan et al. [54], which captures surfel maps using dense frame-to-model tracking and explicitly handles loop closures using non-rigid warping. In contrast, our dynamic de-integration and integration of frames mitigates issues with warping artifacts in rigid structures, and moreover produces a high quality continuous surface. Since our approach does not rely on explict loop closure detection, it scales better to scenarios with many loop closures (c.p. Fig. 7 and 9). We additionally compare to the offline Redwood approach [4], using their rigid variant, see Fig. 8 and 9. Note, we do not compare to their newer non-rigid approach, since it fails on most of our dataset sequences. While their approach takes several hours (2.3h - 13.2h for each of our sequences), we achieve comparable quality and better reconstruction of small-scale detail at real-time rates. Note that Redwood does not take color information into account, thus struggling with sequences that contain fewer geometric features.

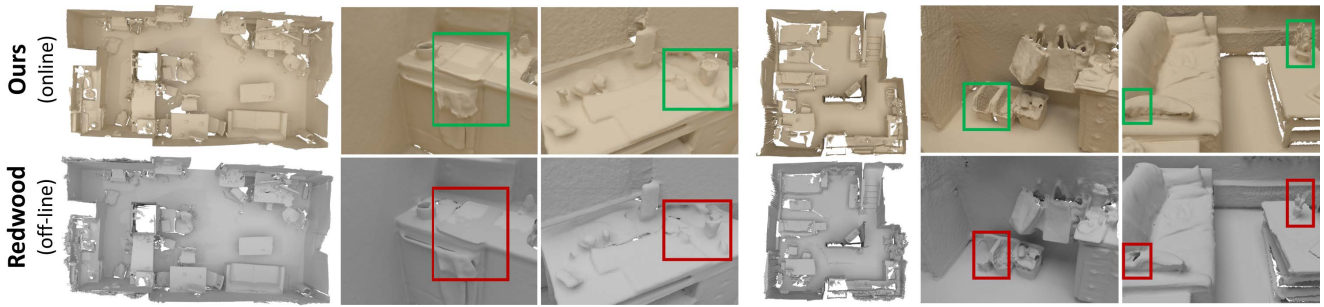

Fig. 8. Our proposed real-time global pose optimization (top) delivers a reconstruction quality on par or even better than the off-line Redwood [4] system (bottom). Note, our reconstructions have more small scale detail.

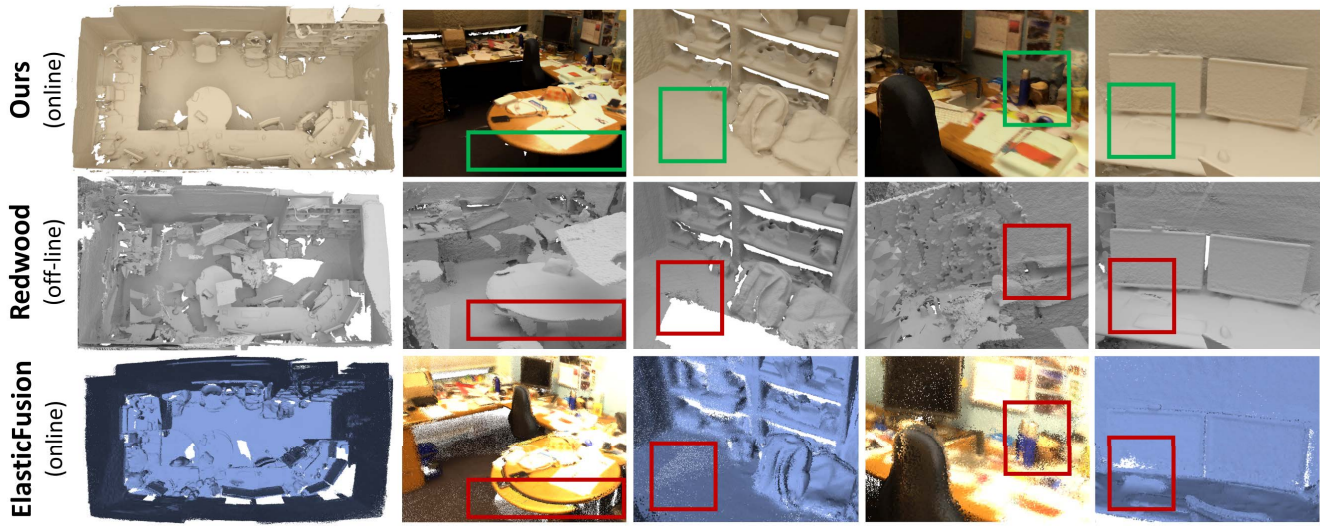

Fig. 9. Our proposed real-time global pose optimization (top) delivers a reconstruction quality on par or even better than the off-line Redwood [4] (middle) and the ElasticFusion [54] (bottom) system. Note that Redwood does not use color information, and was not able to resolve all loop closures in this challenging scan.

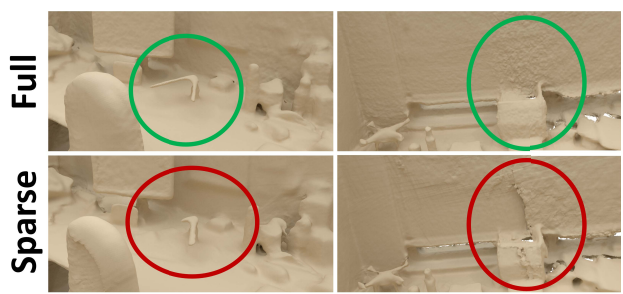

Fig. 10. Comparison of Sparse vs. Dense Alignment: the proposed dense intra- and inter- chunk alignment (top) leads to higher quality reconstructions than only the sparse alignment step (bottom).

*Performance and Convergence.* We measure the performance of our pipeline on an Intel Core i7 3.4GHz CPU (32GB RAM). For compute, we use a combination of a NVIDIA GeForce GTX Titan X and a GTX Titan Black. The Titan X is used for volumetric reconstruction, and the Titan Black for correspondence search and global pose optimization. Our pipeline runs with a framerate well beyond 30Hz (see Fig. 4) for all shown test sequences. Note that the global dense optimization runs in < 500ms at the end of the sequences. After adding a new global keyframe, our approach requires only a few iterations to reach convergence. Fig. 5 shows convergence plots for three of the used test sequences (cf. Fig. 3); the behavior generalizes to all other sequences. We achieve this real-time performance with the combination of our tailored data-parallel Gauss-Newton solver (efficiently handling millions of residuals and solving for over a hundred thousand unknowns), a sparse-to-dense strategy enabling convergence in only a few iterations, and a local-to-global strategy which efficiently decomposes the problem. Note that recent work provides detailed intuition why hand-crafted optimizers outperform existing, general solver libraries [6].

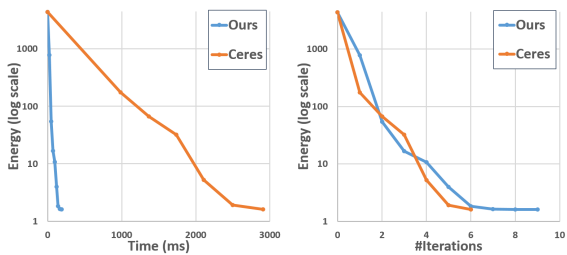

Fig. 11. Performance comparison of our tailored GPU-based solver to Ceres [1]. Both solvers are evaluated over the sparse energy term for 101 keyframes, involving 600 variables and 16339 residuals, with poses initialized to the identity.

Additionally, we evaluate the performance of our tailored GPU-based solver against the widely-used, CPU-based Ceres solver [1]. Fig. 11 shows the performance of both solvers for the sparse energy over 101 keyframes, comprising 600 variables and 16339 residuals, with poses initialized to the identity. Note that this behavior is representative of other sparse energy solves. For Ceres, we use the default Levenberg-Marquardt with a sparse normal Cholesky linear solver (the fastest of the linear solver options for this problem). While our solver takes a couple more iterations to converge without the Levenberg-Marquardt damping strategy, it still runs ≈ 20 times faster than Ceres while converging to the same energy minimum.

Table 1. Memory consumption (GB) for the captured sequences.

| | GPU | | | | | | CPU |
|---|---|---|---|---|---|---|---|
| | | | 1cm | | 4mm | | |
| | Opt-d | Opt-s | Rec | Σ | Rec | Σ | |
| Apt 0 | 1.4 | 0.031 | 0.5 | 1.9 | 3.9 | 5.3 | 20.0 |
| Apt 1 | 1.4 | 0.031 | 0.4 | 1.8 | 3.2 | 4.6 | 20.1 |
| Apt 2 | 0.6 | 0.012 | 0.7 | 1.4 | 6.0 | 6.7 | 9.3 |
| Copyroom | 0.7 | 0.016 | 0.3 | 1.1 | 1.8 | 2.6 | 10.5 |
| Office 0 | 1.0 | 0.021 | 0.4 | 1.4 | 2.5 | 3.5 | 14.4 |
| Office 1 | 0.9 | 0.024 | 0.4 | 1.4 | 2.9 | 3.9 | 13.4 |
| Office 2 | 0.6 | 0.011 | 0.4 | 1.0 | 3.0 | 3.6 | 8.2 |
| Office 3 | 0.6 | 0.011 | 0.4 | 1.0 | 2.7 | 3.3 | 8.9 |

*Memory Consumption.* We evaluate the memory consumption of our globally consistent reconstruction approach on our eight captured sequences, see Tab. 1. The most significant required memory resides in RAM (CPU), i.e., 20GB for Apt 0. It stores all RGB-D frames and depends linearly on the length of the sequence, see Tab. 5. The required device memory (GPU) is much smaller, e.g., 5.3GB (4mm voxels) and 1.9GB (1cm voxels) for the same sequence. This is well within the limits of modern graphics cards (12 GB for GTX Titan X). We also give the amount of memory required to store and manage the TSDF (Rec) and to run the camera pose optimization, both for the sparse term (Opt-s) and the dense term (Opt-d). The footprint for storing the SIFT keypoints and correspondences (included in Opt(s)) is negligibly small; i.e, 31mb for Apt 0. The longest reconstructed sequence (home_at_scan1_2013_jan_1) is part of the SUN3D dataset [57], consisting of 14785 frames (≈ 8.2 minutes scan time @30Hz). This sequence has a CPU memory footprint of 34.7GB and requires 7.3GB of GPU memory (4mm voxels) for tracking and reconstruction.

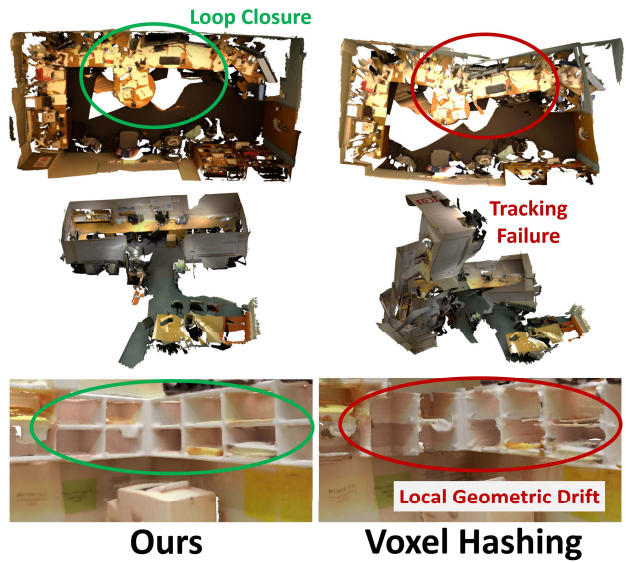

Fig. 12. Comparison to the VoxelHashing approach of Nießner et al. [37]: in contrast to the frame-to-model tracking of VoxelHashing, our novel global pose optimization implicitly handles loop closure (top), robustly detects and recovers from tracking failures (middle), and greatly reduces local geometric drift (bottom).

*Recovery from Tracking Failure.* If a new keyframe cannot be aligned successfully, we assume tracking is lost and do not integrate surface measurements. An example scanning sequence is shown in Fig. 6. To indicate tracking failure, the reconstruction is shown with a gray overlay. Based on this cue, the user is able to recover the method by moving back to a previously scanned area. Note that there is no temporal nor spatial coherence required, as our method globally matches new frames against all existing data. Thus, scanning may be interrupted, and continued at a completely different location.

*Loop Closure Detection and Handling.* Our global pose optimization approach detects and handles loop closures transparently (see Fig. 13), since the volumetric scene representation is continuously updated to match the stream of computed pose estimates. This allows incrementally fixing loop closures over time by means of *integration* and *de-integration* of surface measurements.

*Precision and Recall of Loop Closures.* Tab. 2 gives the precision (i.e., the percentage of correct chunk pair correspondence detections from the set of established correspondences) and recall (i.e., the percentage of detected chunk pair correspondences from the set of ground truth correspondences), on the loop closure set of the augmented ICL-NUIM dataset. A chunk pair correspondence is determined to be in the ground truth set if their geometry overlaps by ≥ 30% according to the ground truth trajectory, and a proposed chunk pair correspondence is determined to be correct if it lies in the ground truth set with reprojection error less than 0.2m, following [4]. We show our registration performance after running the SIFT matcher (Sift Raw), our correspondence filters described in Sec. 4.1.1 – the Key Point Correspondence Filter (SIFT + KF) and the Surface Area and Dense Verification (SIFT + Verify) –, and the final result after the optimization residual pruning described in Sec. 4.4.1 (Opt). As can be seen, all steps of the globally consistent camera tracking increase precision while maintaining sufficient recall.

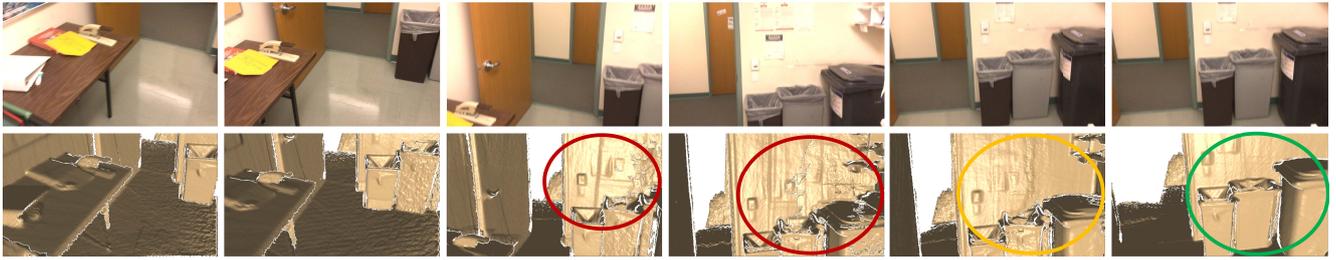

Fig. 13. Global pose optimization robustly detects and resolves loop closure. Note, while data is first integrated at slightly wrong locations, the volumetric representation improves over time as soon as better pose estimates are available.

Table 2. Loop closure precision and recall on the synthetic augmented ICL-NUIM Dataset [4].

|          |               | Sift Raw | Sift + KF | Sift + Verify | Opt  |
|----------|---------------|----------|-----------|---------------|------|
| Living 1 | Precision (%) | 27.0     | 98.0      | 98.2          | 100  |
|          | Recall (%)    | 47.5     | 40.3      | 39.5          | 39.3 |
| Living 2 | Precision (%) | 25.3     | 92.1      | 92.4          | 100  |
|          | Recall (%)    | 49.3     | 47.4      | 45.9          | 45.7 |
| Office 1 | Precision (%) | 14.1     | 97.7      | 99.6          | 100  |
|          | Recall (%)    | 49.1     | 48.7      | 48.0          | 47.7 |
| Office 2 | Precision (%) | 10.9     | 90.2      | 96.2          | 100  |
|          | Recall (%)    | 46.0     | 42.4      | 42.1          | 42.0 |

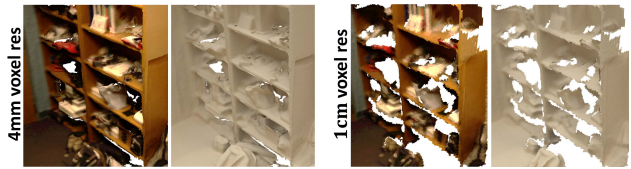

Fig. 14. Comparison of different voxel resolutions: 4mm voxel resolution (left) leads to higher-fidelity reconstructions than the coarser 1cm resolution (right). Note the generally sharper texture and the more refined geometry in case of 4mm voxels.

*Dense Tracking and Voxel Resolution.* In Fig. 10, we evaluate the influence of the dense tracking component of our energy function. While globally drift-free reconstructions can be obtained by sparse tracking only, the dense alignment term leads to more refined local results. The impact of voxel resolution on reconstruction quality is shown in Fig. 14. As a default, we use a voxel resolution of 4mm for all reconstructions. While 1cm voxels reduce memory consumption, the quality of the reconstruction is slightly impaired.

*Quantitative Comparison.* We quantitatively evaluate our approach on independent benchmark data and compare against state-of-the-art online (DVO-SLAM [23], RGB-D SLAM [8], MRSMap [46], Kintinuous [51], VoxelHashing [36, 37], ElasticFusion [54]) and offline systems (Submap Bundle Adjustment [29], Redwood [4]). Note that for Redwood, we show results for the rigid variant, which produced better camera tracking results. We first evaluate our approach on the ICL-NUIM dataset of Handa et al. [17], which provides ground truth camera poses for several scans of a synthetic environment. Table 3 shows our trajectory estimation performance, measured with absolute trajectory error (ATE), on the four living room scenes (including synthetic noise), for which we out-perform existing state-of-the-art online and offline systems. Additionally, in Table 4 we evaluate our approach on the RGB-D benchmark of Sturm et al [47]. This benchmark provides ground truth camera pose estimates for hand-held Kinect sequences using a calibrated motion capture system. For these sequences, which only cover small scenes and simple camera trajectories, our results are on par with or better than the existing state of the art. Note that our own sequences have a larger spatial extent and are much more challenging, with faster motion and many more loop closures.

For these datasets (Tables 3 and 4), the Redwood system, which relies solely on geometric registration, suffers from the relative lack of varying views in the camera trajectories. In particular, fr3/nst is a textured wall, which cannot be registered with a geometric-only method. On both these datasets, we also quantitatively validate the relevance of our design decisions. While online alignment based on sparse features only (Ours (s)) achieves reasonable results, using dense matching only in per chunk alignment further increases accuracy (Ours (sd)). Our full sparse and dense matching approach on both local and global level leads to the highest accuracy.

*Parameters.* While we report default parameters for the Structure Sensor, other RGB-D sensors maintain different noise characteristics, and we vary several parameters accordingly. For significant depth noise, we allow dense verification and residual pruning to be more lax, so as to not acquire too many false negatives. That is, for Kinect data we have a dense reprojection threshold of 0.3m and prune residuals > 0.16m, and for the (noisy) synthetic ICL-NUIM data we

Table 3. ATE RMSE on the synthetic ICL-NUIM dataset by [17].

|  | kt0 | kt1 | kt2 | kt3 |
|---|---|---|---|---|
| DVO SLAM | 10.4cm | 2.9cm | 19.1cm | 15.2cm |
| RGB-D SLAM | 2.6cm | 0.8cm | 1.8cm | 43.3cm |
| MRSMap | 20.4cm | 22.8cm | 18.9cm | 109cm |
| Kintinuous | 7.2cm | 0.5cm | 1.0cm | 35.5cm |
| VoxelHashing | 1.4cm | **0.4cm** | 1.8cm | 12.0cm |
| Elastic Fusion | 0.9cm | 0.9cm | 1.4cm | 10.6cm |
| Redwood (rigid) | 25.6cm | 3.0cm | 3.3cm | 6.1cm |
| Ours (s) | 0.9cm | 1.2cm | 1.3cm | 1.3cm |
| Ours (sd) | 0.8cm | 0.5cm | 1.1cm | 1.2cm |
| Ours | **0.6cm** | **0.4cm** | **0.6cm** | **1.1cm** |

Note that unlike the other methods, Redwood does not use color information and runs offline. For our approach, we also provide results for sparse-only (s) as well as sparse and local dense only (sd).

Table 4. ATE RMSE on the TUM RGB-D dataset by [47].

|  | fr1/desk | fr2/xyz | fr3/office | fr3/nst |
|---|---|---|---|---|
| DVO SLAM | 2.1cm | 1.8cm | 3.5cm | 1.8cm |
| RGB-D SLAM | 2.3cm | **0.8cm** | 3.2cm | 1.7cm |
| MRSMap | 4.3cm | 2.0cm | 4.2cm | 201.8cm |
| Kintinuous | 3.7cm | 2.9cm | 3.0cm | 3.1cm |
| VoxelHashing | 2.3cm | 2.2cm | 2.3cm | 8.7cm |
| Elastic Fusion | 2.0cm | 1.1cm | **1.7cm** | 1.6cm |
| LSD-SLAM | - | 1.5cm | - | - |
| Submap BA | 2.2cm | - | 3.5cm | - |
| Redwood (rigid) | 2.7cm | 9.1cm | 3.0cm | 192.9cm |
| Ours (s) | 1.9cm | 1.4cm | 2.9cm | 1.6cm |
| Ours (sd) | 1.7cm | 1.4cm | 2.8cm | 1.4cm |
| Ours | **1.6cm** | 1.1cm | 2.2cm | **1.2cm** |

Note that unlike the other methods listed, Redwood does not use color information and runs offline. For our approach, we also provide results for sparse-only (s) as well as sparse and local dense only (sd).

have a dense reprojection threshold of 0.1m and prune residuals > 0.08m.

*Limitations.* As our tracking is based on sparse key point matching, small local misalignments can occur; e.g., SIFT matches can be off by a few pixels and the depth data associated with a keypoint may be inaccurate due to sensor noise. While we solve for optimal keypoint positions from the inter-chunk optimization, small mismatches between global keypoints can still be propagated within the global optimization, leading to local misalignments. Ideally, we would treat the locations of the global keypoints as unknowns to optimize for. Unfortunately, this would involve significant computational effort, which (currently) seems to exceed even the computational budget of offline approaches. Another limitation is that we currently run our method on two GPUs. Fortunately, we can easily stream the data to and from an iPad with live visual feedback, on both the desktop and mobile device, thus making scanning fun and convenient. With our current hardware configurations, we are limited to scans of up to 25,000 input RGB-D frames. This corresponds to about 14 minutes of continuous scanning, assuming 30Hz input – although many RGB-D sensors have a lower frame rate which allows for longer sessions. In order to allow for longer sequences, we would need more than two hierarchy levels to perform the optimization in real time. We could also imagine spatial clustering – e.g., into separate rooms – and split up the optimization tasks accordingly.

## 7 ADDITIONAL EVALUATION

Table 5. Dataset overview.

|  | #Frames | Trajectory Length |
|---|---|---|
| Apt 0 | 8560 | 89.4m |
| Apt 1 | 8580 | 91.8m |
| Apt 2 | 3989 | 87.5m |
| Copyroom | 4480 | 24.4m |
| Office 0 | 6159 | 52.8m |
| Office 1 | 5730 | 51.7m |
| Office 2 | 3500 | 36.3m |
| Office 3 | 3820 | 66.8m |

In order to capture scans at high completeness, the camera is moved in long and complex trajectories.

### 7.1 Additional Qualitative Results

Reconstructed models for the eight scenes in our dataset are publicly available [3]. While our method and ElasticFusion run at real-time rates, Redwood runs offline, taking 8.6 hours for Apt0, 13.2 hours for Apt1, 4 hours for Copyroom, 7.7 hours for Office1, 2.6 hours for Office2, and 3.5 hours for Office3. The relocalization (due to sensor occlusion) in the sequence Apt 2 cannot be handled by state-of-the-art methods such as ElasticFusion and Redwood. Redwood is also a geometry-only approach that does not use the RGB channels. Note that the lack of ElasticFusion results on some sequences is due to the occasional frame jump in our wifi streaming setup, which dense frame-to-model methods cannot handle.

We additionally evaluate our method on the SUN3D dataset [57], which contains a variety of indoor scenes captured with an Asus Xtion sensor. Fig. 15 shows reconstruction results for several large, complex scenes, using the offline SUN3Dsfm bundle adjustment system as well as our approach. Note that our approach produces better global structure while maintaining local detail at real-time rates. The SUN3D dataset also contains eight scenes which contain manual object-correspondence annotations in order to guide their reconstructions; we show reconstruction results using our method (without annotation information) on these scenes in Fig. 16.

In addition, we have reconstructed all 464 scenes from the NYU2 dataset [43], which contains a variety of indoor scenes recorded by a Kinect. Several reconstruction results are shown in Fig. 17.

### 7.2 Additional Quantitative Results

The ICL-NUIM dataset of Handa et al. [17] also provides the ground truth 3D model used to generate the virtually scanned sequences. In addition to the camera tracking evaluation provided in Section 6 of the paper, we evaluate surface reconstruction accuracy (mean distance of the model to the ground truth surface) for the living room model in Table 6.

---

[3]http://www.graphics.stanford.edu/projects/bundlefusion/

| | SUN3Dsfm | Ours |
|---|---|---|
| **home_at_scan1:**<br><br>14785 frames<br>102.6m trajectory | 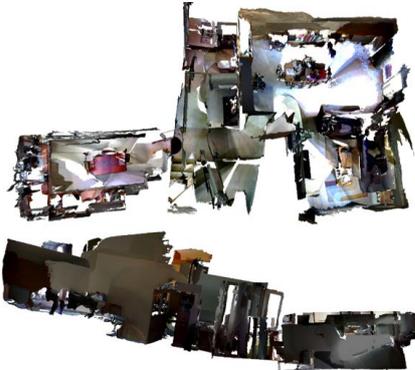 | 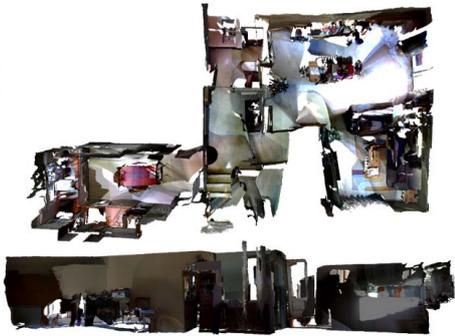 |
| **scan1:**<br><br>9526 frames<br>109.4m trajectory | 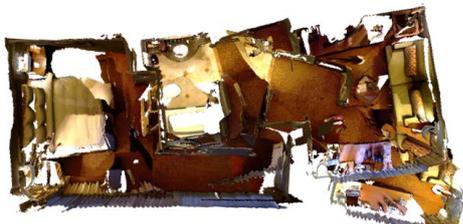 | 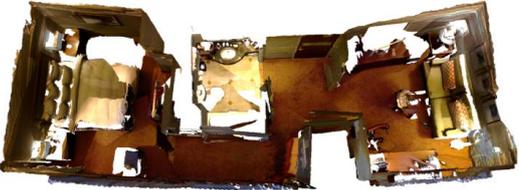 |
| **scan3:**<br><br>10207 frames<br>70.1m trajectory | 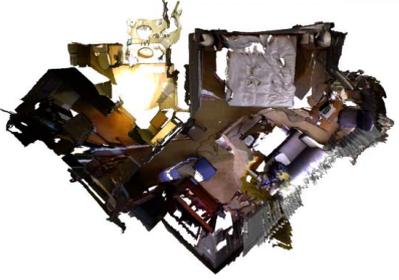 | 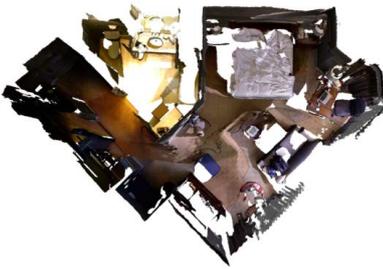 |
| **nips_4:**<br><br>6269 frames<br>58.1m trajectory | 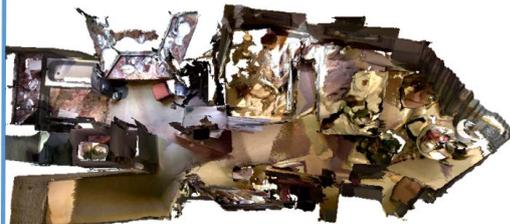 | 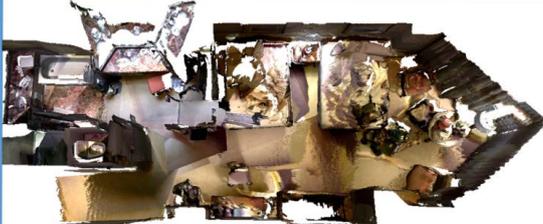 |
| **hotel_umd_1:**<br><br>5693 frames<br>49.6m trajectory | 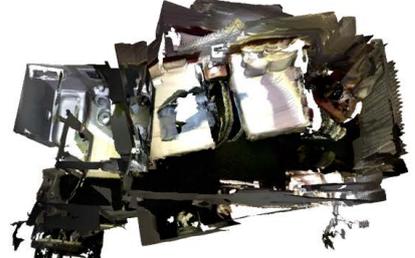 | 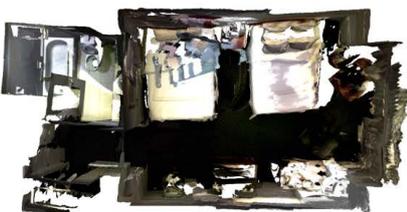 |

Fig. 15. Reconstruction results on scenes from the SUN3D dataset [57], using SUN3Dsfm and our approach.

# Reconstruction results on SUN3D annotated scenes

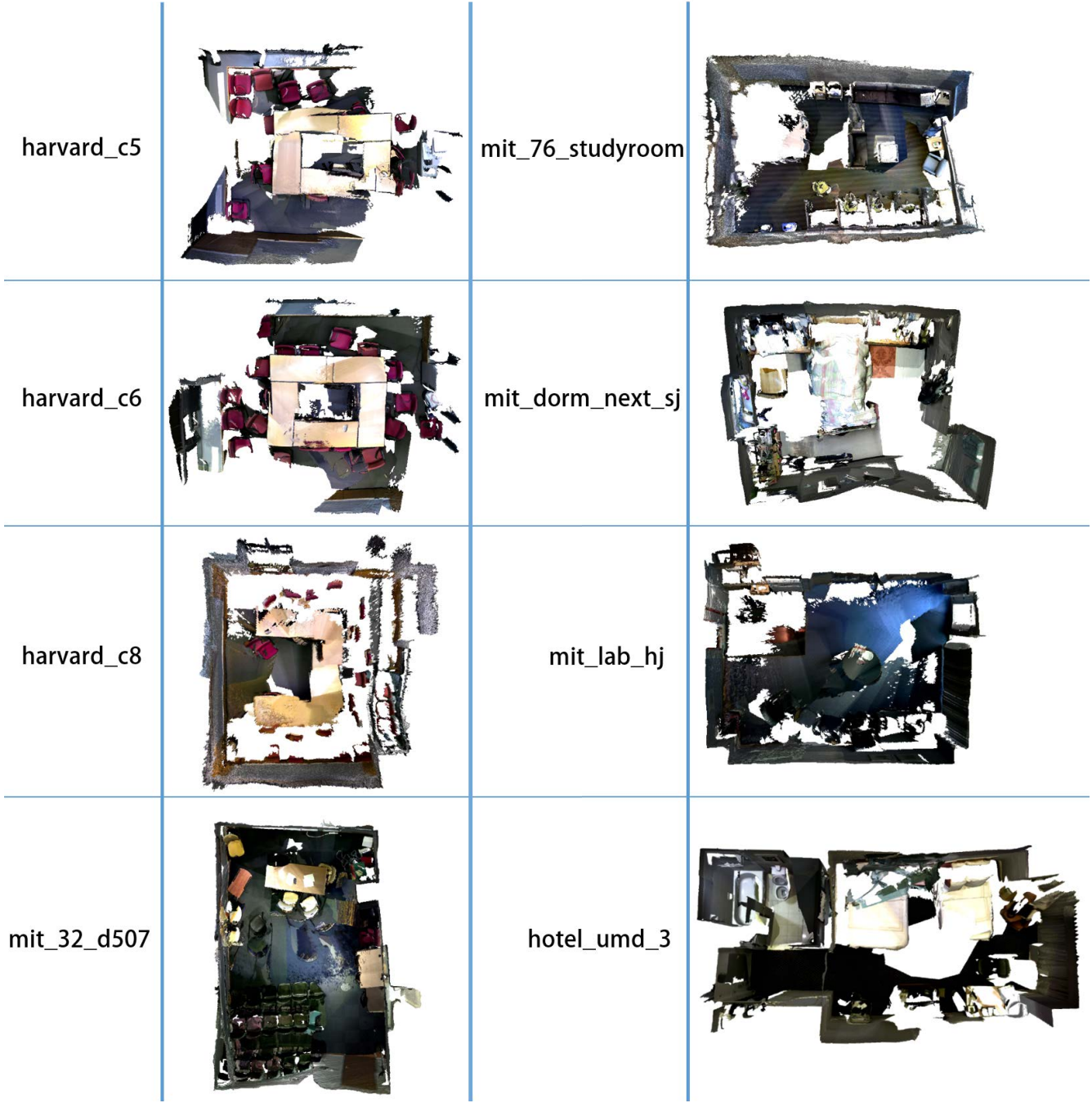

Fig. 16. Reconstruction results on eight scenes from the SUN3D dataset [57], chosen from the *List of Annotated Scenes* (our method is fully automated and does not use any annotations).

## Reconstructions of NYU2 Scenes

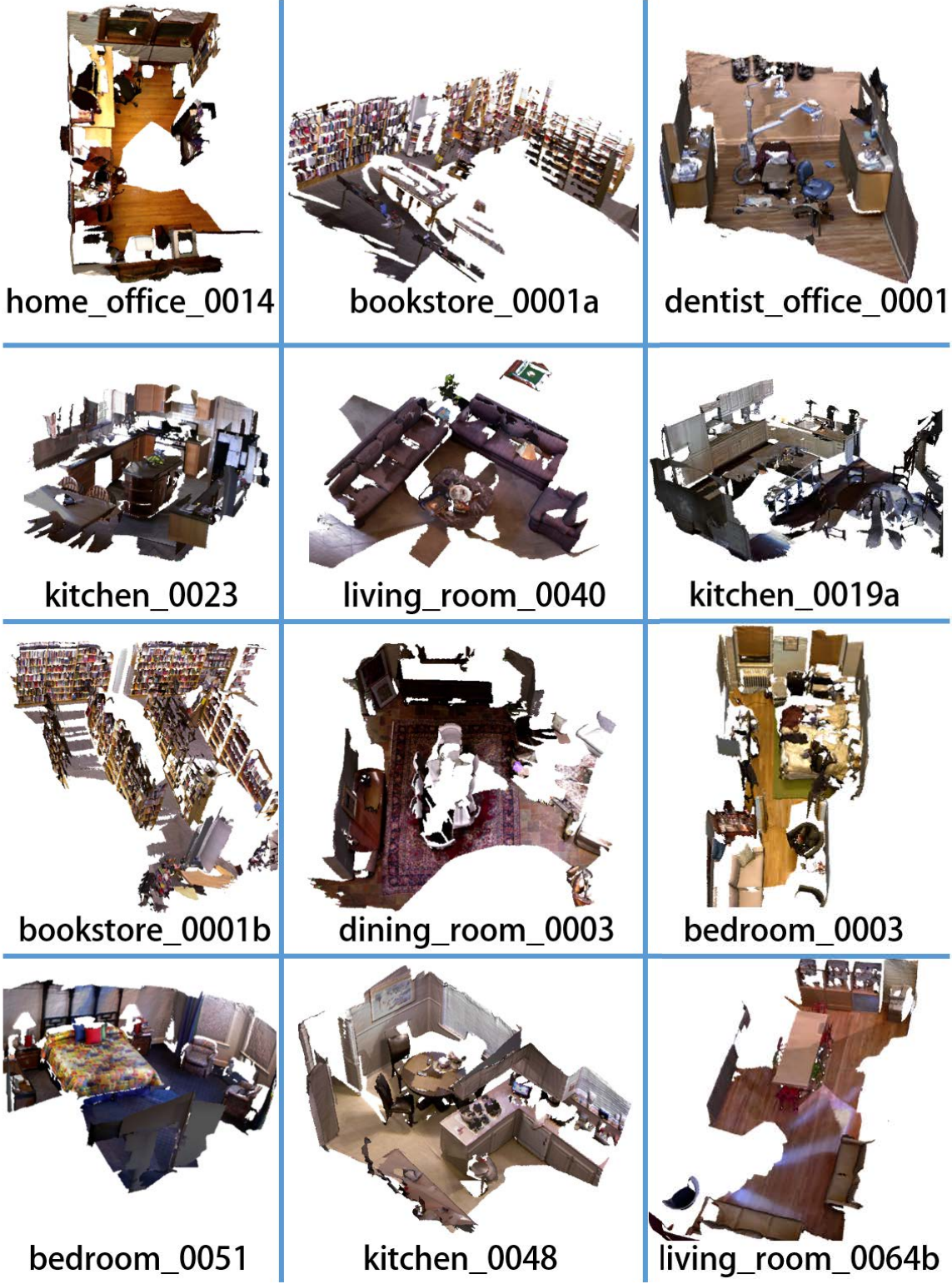

Fig. 17. Reconstructions from the NYU2 dataset [43].

Additionally, we further evaluate our camera tracking on the augmented ICL-NUIM dataset of [4], which comprises synthetic scans of two virtual scenes, a living room and an office, from the original ICL-NUIM data. In contrast to the original ICL-NUIM, these scans have longer trajectories with more loop closures. Table 7 shows our trajectory estimation performance on this dataset (with synthetic sensor noise, using the reported camera intrinsic parameters), which is on par with or better than existing state of the art. Although the camera trajectories are complex, the additional loop closures help maintain stability (as frames find matches which are not neighbors, mitigating tracking drift), aiding our performance in all scenes except Office 1. In this case, our method has difficulty closing the loop, as part of it covers a wall with little to no color features.

In Table 8, we show the number of frames registered for each our captured sequences, as well as for the augmented ICL-NUIM and various SUN3D sequences. Our method registers the vast majority of frames in these sequences, only dropping frames when they fail to pass our correspondence filters, which err on the conservative side. Many of the unregistered frames listed contain sensor occlusions (for the Apt 2 relocalizations) or untextured walls, indicating that our correspondence and frame filtering finds a good balance between discarding potentially bad matches and retaining good matches to maintain stable tracking.

Table 6. Surface reconstruction accuracy on the synthetic ICL-NUIM dataset by [17].

|  | kt0 | kt1 | kt2 | kt3 |
|---|---|---|---|---|
| DVO SLAM | 3.2cm | 6.1cm | 11.9cm | 5.3cm |
| RGB-D SLAM | 4.4cm | 3.2cm | 3.1cm | 16.7cm |
| MRSMap | 6.1cm | 14.0cm | 9.8cm | 24.8cm |
| Kintinuous | 1.1cm | 0.8cm | 0.9cm | 24.8cm |
| Elastic Fusion | 0.7cm | 0.7cm | 0.8cm | 2.8cm |
| Redwood (rigid) | 2.0cm | 2.0cm | 1.3cm | 2.2cm |
| Ours | **0.5cm** | **0.6cm** | **0.7cm** | **0.8cm** |

Mean distance of each reconstructed model to the ground truth surface. Note that unlike the other methods listed, Redwood does not use color information and runs offline.

Table 7. ATE RMSE on the synthetic augmented ICL-NUIM Dataset by [4].

|  | Living room 1 | Living room 2 | Office 1 | Office 2 |
|---|---|---|---|---|
| Kintinuous | 27cm | 28cm | 19cm | 26cm |
| DVO SLAM | 102cm | 14cm | 11cm | 11cm |
| SUN3D SfM | 21cm | 23cm | 24cm | 12cm |
| Redwood | 10cm | 13cm | **6cm** | 7cm |
| Ours | **0.6cm** | **0.5cm** | 15.3cm | **1.4cm** |

Note that unlike the other methods listed, Redwood does not use color information and runs offline.

## 7.3 SIFT Performance

We provide an additional performance analysis of our GPU-based SIFT detection and matching strategy, see Table 9. Note that for a $1296 \times 968$ image (another Structure sensor color resolution), SIFT detection time increases slightly to $\approx$ 6.4ms. We detect $\sim$ 150 features per frame, and $\sim$ 250 per keyframe, for all sequences.

Table 8. Frame Registration and Missed frames.

|  | #Frames | #Unregistered Frames |
|---|---|---|
| Apt 0 | 8560 | 0 |
| Apt 1 | 8580 | 82 |
| Apt 2 | 3989 | 115 |
| Copyroom | 4480 | 0 |
| Office 0 | 6159 | 105 |
| Office 1 | 5730 | 1 |
| Office 2 | 3500 | 0 |
| Office 3 | 3820 | 42 |
| SUN3D (avg) | 5100 | 6 |
| Livingroom 1 (A-ICL) | 2870 | 1 |
| Livingroom 2 (A-ICL) | 2350 | 0 |
| Office 1 (A-ICL) | 2690 | 94 |
| Office 2 (A-ICL) | 2538 | 0 |

Frames registered by our method on various sequences. The unregistered frames are those which did not find sufficient sparse matches; i.e., untextured walls or frames in which the depth sensor was occluded. For instance, in *Apt 2*, we occlude the sensor with our hand to demonstrate our relocalization ability, which leads to a higher unregistered frame count. Note that the SUN3D frame registration is reported for the average of the scenes shown in Figs. 15-16, and that A-ICL refers to the Augmented ICL-NUIM dataset.

Table 9. SIFT performance for a $640 \times 480$ image.

| #Features | Time Detect (ms) | Time Match (ms) |
|---|---|---|
| 150 | 3.8 | 0.04 |
| 250 | 4.2 | 0.07 |
| 1000 | 5.8 | 0.42 |

Detection time (including descriptor computation) is reported per frame, and match time per image pair (parallelized). On all sequences run, we detect about 150 features per frame, and about 250 per keyframe.

## 8 CONCLUSION

We have presented a novel online real-time 3D reconstruction approach that provides robust tracking and implicitly solves the loop closure problem by globally optimizing the trajectory for every captured frame. To this end, we combine online SIFT feature extraction, matching, and pruning with a novel parallel non-linear pose optimization framework, over both sparse features as well as dense correspondences, enabling the solution of the global alignment problem at real-time rates. The continuously changing stream of optimized pose estimates is monitored and the reconstruction is updated through dynamic integration and de-integration. The capabilities of the proposed approach have been demonstrated on several large-scale 3D reconstructions with reconstruction quality and completeness that was previously only possible with offline approaches and tedious capture sessions. We believe online global pose alignment will pave the way for many new and interesting applications. Global accurate tracking is the foundation for immersive AR/VR applications and makes online hand-held 3D reconstruction applicable to scenarios that require high-fidelity tracking.

## ACKNOWLEDGMENTS

We would like to thank Thomas Whelan for his help with Elastic-Fusion, and Sungjoon Choi for his advice on the Redwood system.

This work was funded by the Max Planck Center for Visual Computing and Communications, the ERC Starting Grant 335545 CapReal, and a Stanford Graduate Fellowship. We are grateful for hardware donations from NVIDIA Corporation and Occipital.